\documentclass[fleqn,usenatbib]{mnras}

\usepackage{newtxtext,newtxmath}

\usepackage[T1]{fontenc}

\DeclareRobustCommand{\VAN}[3]{#2}
\let\VANthebibliography\thebibliography
\def\thebibliography{\DeclareRobustCommand{\VAN}[3]{##3}\VANthebibliography}

\usepackage{graphicx}	
\usepackage{amsmath}	
\usepackage{xcolor}
\usepackage{ulem}

\newcommand{\chandra}{\textit{Chandra}}
\newcommand{\rxte}{\textit{RXTE~}}

\newcommand{\xmm}{\textit{XMM-Newton}}
\newcommand{\sax}{\textit{BeppoSAX~}}
\newcommand{\fermi}{\textit{Fermi}}
\newcommand{\swift}{\textit{Swift}}
\newcommand{\integral}{\textit{INTEGRAL}}
\newcommand{\nustar}{\textit{NuSTAR}}

\newcommand{\src}{MXB 0656-072}
\newcommand{\code}{\texttt}
\newcommand{\srcc}{MAXI J1409-619}

\title[]{Quiet, but not silent: Uncovering quiescent state properties of two transient High Mass X-ray binaries}

\author[Raman et al.]{
Gayathri Raman,$^{1}$\thanks{E-mail: gzr5209@psu.edu}
Varun,$^{2}$
Pragati Pradhan,$^{3}$
Jamie Kennea$^{1}$
\\
$^{1}$Department of Astronomy and Astrophysics, The Pennsylvania State University, 525 Davey Lab, University Park, PA 16802, USA\\
$^{2}$Aryabhatta Research Institute of Observational Sciences (ARIES), Manora Peak, Nainital-263001, Uttarakhand, India\\
$^{3}$Embry-Riddle Aeronautical University, Department of Physics and Astronomy, 3700 Willow Creek Road, Prescott,
AZ 86301, USA
}

\date{Accepted XXX. Received YYY; in original form ZZZ}

\begin{document}
\label{firstpage}
\pagerange{\pageref{firstpage}--\pageref{lastpage}}
\maketitle

\begin{abstract}
We present the first set of broadband spectral and timing studies of two transient X-ray pulsars, \src\ and \srcc\,  using \nustar ~observations conducted during quiescence. Despite being captured at one of their lowest luminosity states, both these targets show signs of ongoing low-level accretion. Results from the time-averaged spectral analysis indicate for the first time, the presence of a strong soft power law component along with thermal emission from the neutron star hot spots. For both targets, the quiescent thermal X-ray emission is consistent with the deep crustal heating model. In \src, we do not detect any pulsations or indications of a cyclotron line during quiescence. However, in \srcc ~we detect strong pulsations at 502~s with a pulsed fraction of $\sim$66\%, which adds this pulsar to the list of a handful of quiescent-state pulsating systems.
\end{abstract}




\section{Introduction}

Accretion powered X-ray pulsars exhibit a broad dynamical range of X-ray luminosities ranging from 10$^{32-34}$~ergs~s$^{-1}$ during the quiescent phase all the way up to 10$^{37-38}$~ergs~s$^{-1}$ during outbursts, making them ideal laboratories to probe accretion flows in various accretion regimes. While there have been numerous studies of these systems during their highest luminosity states (see for example, \citealt{Bachetti2014,Doroshenko2020,Raman2021}), it is only with the advent of more sensitive missions like \xmm,~ \chandra~ and \nustar, that has it become possible to explore the low accretion rate regime of these sources in greater detail (for example, \citealt{Campana2002,Roth2013,Elshamouty2016,Tsygankov2017quies,Escorial2020}). The advantage of quiescent state observations is that they offer a direct view of the emission from the neutron star (NS) atmosphere and hotspots without having to deal with additional complexities that come with increased accretion such as the presence of an accretion column and interactions of X-ray photons with matter in the vicinity of these columns. Moreover, transient pulsars spend a considerable amount of time in quiescence during which we anticipate the observation of cooling effects in the neutron star (NS) crust. These effects are directly influenced by the pulsar's outburst history. Such cooling processes can significantly contribute to X-ray emission and/or variability, making it crucial to enhance our understanding of these low accretion states through careful observation.

As the post-outburst luminosity decreases, the NS spin (P$_{\rm{spin}}$) and the surface magnetic field become crucial parameters in determining the system evolution \citep{Escorial2020}. In particular, for fast spinning pulsars at lower accretion states, their magnetospheric radius expands to beyond their co-rotation radius. This effect centrifugally inhibits the infall of incoming material and further expels material outwards. This is most commonly referred to as the `propeller effect' \citep{IllSun1975}. More recent works have suggested some improvement in the standard `centrifugal barrier' picture \citep{Dangelo2011,DAngelo2012}. According to these works, the mass transferred from a binary companion onto a spinning magnetosphere might be accreted cyclically rather than being propelled out. This takes place via the formation of a `Dead disc' which is coupled with the NS magnetosphere. Observations have indicated that residual X-ray emission is observed in several Be X-ray pulsars at very low luminosities \citep{Tsygankov2017quies}. There has also been evidence of pulsations being detected in several objects such as A0535+262 \citep{Roth2013,Doroshenko2014, Tsygankov2017quies}, 4U 1145-619 \citep{Rutledge2007}, 1A 1118-615 \citep{Rutledge2007}, among others, indicating that matter might still be reaching the NS surface \citep{Escorial2020}. \\
The powering mechanism of quiescent state (below or close to the propeller luminosity) X-ray emission is still highly uncertain. \citet{Campana2001} proposed that there could be matter leaking via the magnetospheric centrifugal barrier and contributing to the observed emission. In particular, for slow rotating systems (P$_{\rm{spin}} \sim$ 10s of seconds), it has been shown that the required matter leakage can be achieved via quasi-stable accretion from a `cold' recombined disk \citep{Tsygankov2017cold}. Rapidly rotating systems will be unable to achieve a cold disk because of the onset of the propeller regime at fairly higher mass accretion rates. Another proposed origin for residual X-ray emission from quiescent pulsars is the `deep crustal heating' model \citep{Brown1998}. Freshly accreted matter compresses the NS crust thus inducing nuclear reactions that power the thermal emission observed during quiescence. This mechanism has been successful in explaining the quiescent state emission for several X-ray binaries hosting NS with low magnetic field strengths (for eg. LMXBs, \citealt{WijDegenaar2017}). In particular, for systems with no recent outburst episodes or other signatures of increased activity, it can be safely assumed that the neutron star crust and core are at thermal equilibrium \citep{WijDegenaar2017}. In such cases, we can probe the core temperatures and in turn, examine NS core physics by using the NS surface temperature as a proxy. Additionally, if the NS is highly magnetized, we can further study the effect of the magnetic field on the heating and cooling of NS crusts. 

 The Be X-ray binary transient, \src, was first discovered by \textit{SAS-3} in 1975 at a flux density of 80 mCrab \citep{Clark1975}. It was followed by two more observations at 50 and 70 mCrab in 1976 \citep{Kaluzienski1976} and was initially classified to be a low mass X-ray binary \citep{Liu2001}. After several decades of remaining in quiescence, the source went into an outburst in 2003 for a duration of about 2 months \citep{Rem-Mar2003}. The source was identified as a pulsating object with periodicity at 160.7~s \citep{Morgan2003}.  The optical counterpart was identified as an O9.7 Ve spectral type star using ROSAT PSPS observations \citep{Pakull2003} (later refined and re-classified as a O9.5 Ve star, \citealt{Nespoli2012}). The 2003 outburst was classified as a type-II outburst using \rxte observations. Preliminary \rxte results during that outburst revealed the presence of a cyclotron resonant scattering feature at$\sim$33~keV \citep{Heindl2003}. Further detailed studies of the outburst characteristics were carried out by \citet{McBride2006}, where they found that the width of the Cyclotron Resonant Scattering Feature (CRSF) increased during the decline, while no changes in CRSF properties were observed as a function of pulse phase. With an average pulse period of 160.4$\pm$0.4~s, a spin-up of 0.45~s was observed across the outburst \citep{McBride2006}. Using optical observations, the source was estimated to be located at a distance of 3.9$\pm$0.1~kpc \citep{Pakull2003,McBride2006}. The next set of outbursts took place between November 2007 to November 2008 that was monitored using \integral\ \citep{Kreykenbohm2007}, \rxte\ \citep{Pottschmidt2007} and \swift-BAT \citep{Kennea2007}. Optical observations conducted between 2006-2009, indicated a steadily strengthening H$\alpha$ line, whose equivalent width was strongest just prior to the onset of the 2007 outburst \citep{Yan2007}. This was attributed to an extended circumstellar disk. Moreover, correspondingly fainter UBV magnitudes, during the same 2007 outburst, were indicative of inner disk dilution \citep{Yan2007}. Long term X-ray observations using \swift and \rxte during the period 2006 to 2009 indicated an orbital period of 101.2~days, for the first time \citep{Yan2012}. The HEXTE spectrum obtained during outburst was described using a cutoff power law with a low energy absorption, along with a 6.4~keV Fe line and a CRSF at 30~keV \citep{Yan2012}. Additional X-ray/optical variability studies identified a correlation between the aperiodic variability and spectral parameters, similar to 1A 1118-615 \citep{Nespoli2012}. 


\srcc ~is a transient first detected on board MAXI during outburst in October 2010  \citep{Yamaoka2010ATel2959}. This was localized to a position corresponding to RA=14:08:02.56 and Dec=-61:59:00.3 using \swift-XRT \citep{Kennea2010ATelb}. The lack of an optical counterpart and the presence of a nearby IR source suggested a possible High Mass X-ray Binary nature for the transient \citep{Orlandini2012}. In November 2010, \srcc ~went into its second outburst that was 7 times brighter than its previous event. This triggered the \swift ~Burst Alert Telescope (BAT). Pointed XRT observations of the source in the PC mode  revealed the presence of a 503$\pm$10~s periodicity with a 42\% pulsed fraction, which was associated with the spin period of the pulsar in the system \citep{Kennea2010ATel2962a}. \fermi-GBM observations from December 2010, indicated a refined spin period of 506.95~s with the pulse profile having double peaked shape \citep{CamArr2010ATel}. Interestingly, Quasi Periodic Oscillations (QPOs) were detected in this object at 0.192$\pm$0.006~Hz along with two higher harmonics using \rxte observations \citep{Kaur2010ATel3082}. More recently, \citet{Donmez2020} also reported correlations between the source flux state and the presence and strength of the QPOs and their harmonics. This source and 4U 0115+63 are the only two known accreting X-ray pulsars to exhibit QPO harmonics \citep{Donmez2020}. The source continuum spectrum studied using \rxte  observations during outburst was well described by a cutoff power law with index 1.3 along with either a partially covered absorption model or a reflection model \citep{Donmez2020}. The spectrum also showed a strong 6.5~keV Fe line with no indication of a cyclotron feature \citep{Yamamoto2010ATel3070}. The source was re-discovered with archival \sax observations from 2000 during its low state. The broadband \sax spectrum in the 1.8--100~keV band was best described using an absorbed power law ($\Gamma\sim$0.8) \citep{Orlandini2012}. A cyclotron absorption feature with a fundamental at 44~keV (along with two higher harmonics at 73~keV and 128~keV) was also detected in the quiescent state that allowed for the measurement of the neutron star surface magnetic field of 3.8$\times$10$^{12}$~G \citep{Orlandini2012}.

The two X-ray pulsars analyzed as part of this work have been in their quiescent accretion state for more than a decade. A dedicated observing campaign was conducted using \chandra, \xmm\ and \swift\ to explore the quiescent state of X-ray pulsars \citep{Wij2003cxoprop}. As part of that observing campaign in 2012, \src ~was observed using Chandra, where it displayed a soft thermal spectrum \citep{Tsygankov2017quies}. The source was observed at a flux level of 2$\times$10$^{-12}$~erg~cm$^{-2}$s$^{-1}$, which corresponds to a luminosity of 6.2$\times$10$^{33}$~erg~s$^{-1}$ (assuming a distance of 5.1~kpc obtained using recent Gaia measurements \citealt{Arnason2021}). 
 \srcc ~has been in quiescence since its last outburst in 2010 and has never been studied using current instruments since. In this paper, we study the broadband spectral properties of these two transient X-ray pulsars using \nustar ~observations. In Section 2, we present the methods and observation details, followed by the analysis and results in Section 3. We further summarize our results and discussions in Section 4. 
 

\section{Observations and data reduction}

\begin{figure}
    \centering
    \includegraphics[scale=0.35,angle=0,trim={0.5cm 5.8cm 0cm 6cm},clip]{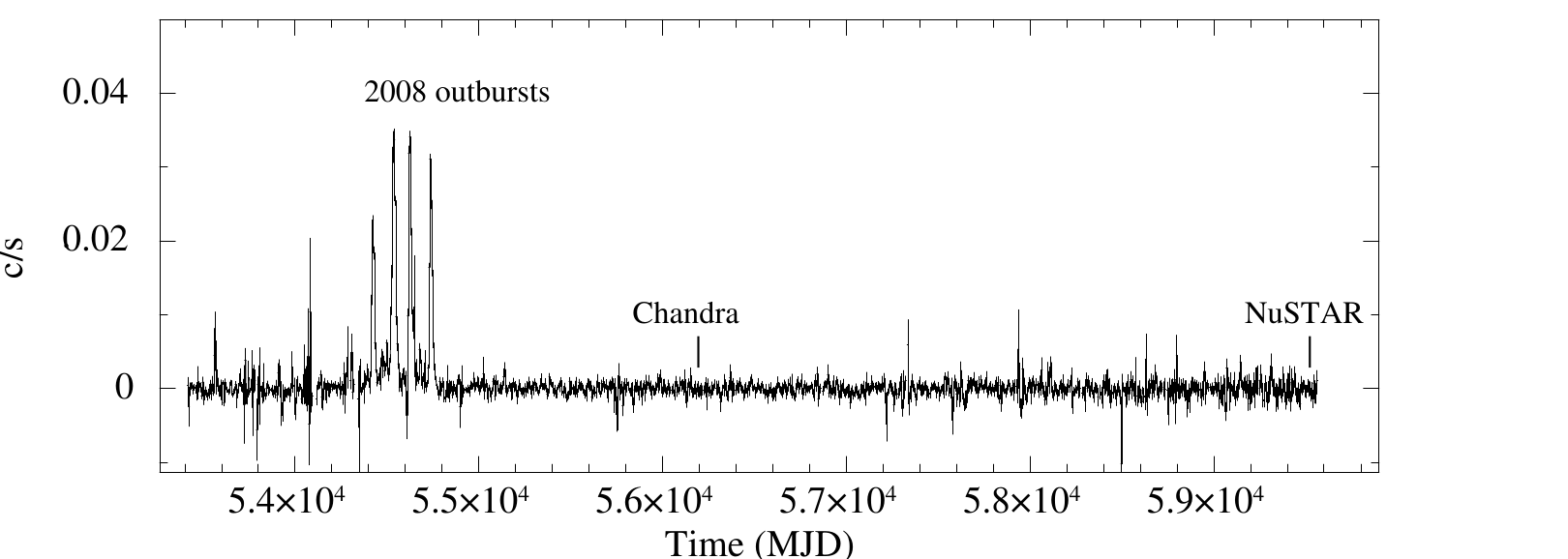}
    
    \includegraphics[scale=0.35,angle=0,trim={0.5cm 5.2cm 0cm 6cm},clip]{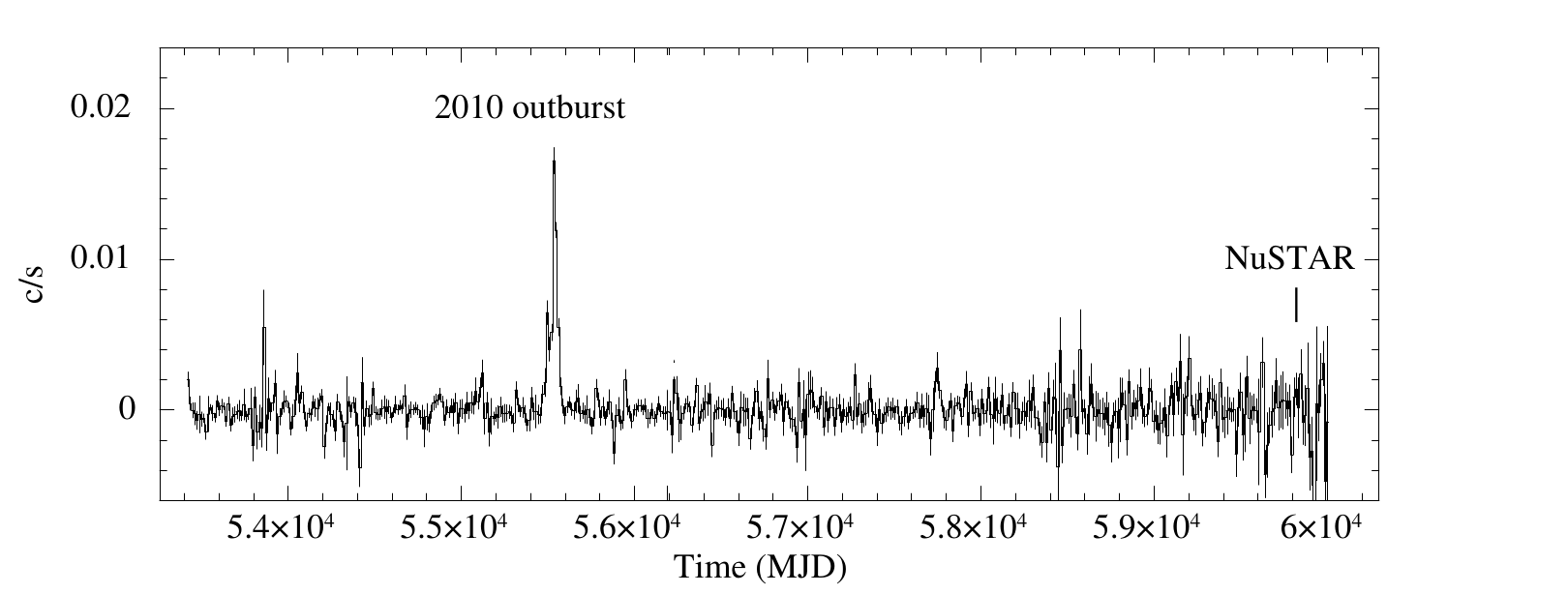}
    \caption{ Figure shown the long term \swift-BAT (15--50~keV) light curves for \src ~(top panel) and \srcc ~(bottom panel) with the \nustar ~and \chandra~ quiescent state observations marked. Both targets have remained in quiescence since their previous major outbursts.}
    \label{fig:bat-lc}
\end{figure}

The Nuclear Spectroscopic Telescope Array (\nustar) is a space-based high energy mission capable of carrying out sensitive X-ray imaging and spectroscopy in the 3--79~keV band. It comprises of two co-aligned identical X-ray focal plane modules FPMA and FPMB, each with a FOV of  13$'\times$13$'$  \citep{Harrison2013}.  It provides a moderate spectral resolution of 400~eV around 10~keV. With its broadband spectral sensitivity, \nustar ~is ideally suited to carry out cyclotron line studies and broadband spectroscopy. 
The \nustar ~observations of \src ~and \srcc ~were carried out in November 2021 and September 2022 with an exposure time of 50~ks and 56~ks, respectively. Details of the observations are given in Table \ref{tab:tab1}. Both sources were observed during the ongoing low accretion states. Figure \ref{fig:bat-lc} shows the long-term \swift-BAT 15-50~keV light curve marking the previous outbursts and the location of the \nustar ~observations used for this work.




\begin{figure*}
    \centering
    \includegraphics[scale=0.75,angle=0,trim={1cm 8.8cm 1cm 7.8cm},clip]{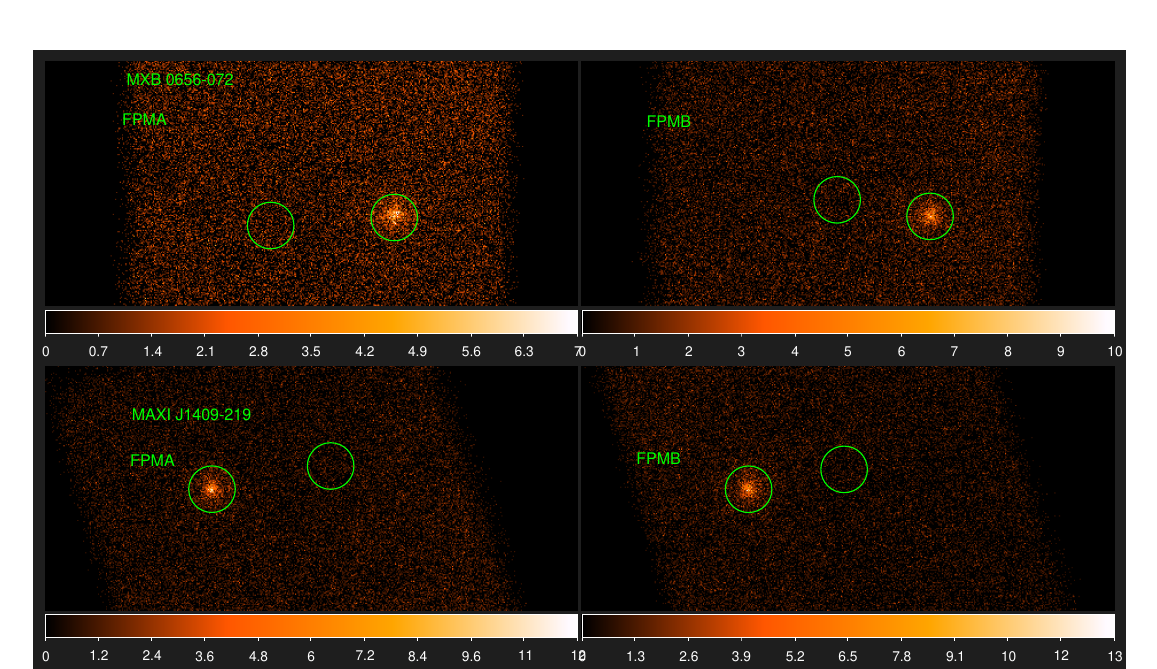}
    \caption{\nustar ~images are shown for \src ~(top) and \srcc ~(bottom) in the two detector modules with the source and background regions marked. }
    \label{fig:my_label}
\end{figure*}

\begin{table*}
    \centering
    \begin{tabular}{|c|c|c|c|c|c|}
    \hline
        Target & Obs ID & T$_{\rm start}$ (UTC)  & MJD & Exp time & Count rate  \\
       & &  & & (ks) & (s$^{-1}$)\\
        \hline
        
        \src & 30701016002 & 2021-11-10 11:02:24 & 59528.46 & 50 & 0.04\\
        \srcc &30801023002 & 2022-09-06 22:56:09 & 59828.95 & 56 & 0.06\\
        
        
         \hline
         
    \end{tabular}
    \caption{\nustar ~observations of \src ~and \srcc ~used in this work. }
    \label{tab:tab1}
\end{table*}

The raw data were processed using the standard \nustar ~data analysis software \code{nustardas} version 2.1.1 along with the CALDB version 20210210. Source photons were extracted from a circular region with radius 50 arc seconds for both targets. The S/N was optimized by collecting background photons from a nearby source-free region, using the standard background extraction procedure. We used the routine \code{NUPIPELINE} to generate the calibrated level 2 event files. The photon arrival times in the source event file were corrected to the solar center barycenter using the target coordinates (in J2000): RA=104.57 degrees and Dec=-7.209 degrees for \src, and RA=212.01 degrees and Dec=-61.98342 degrees for \srcc. The light curves, spectra, and other auxiliary response files were extracted using the task \code{nuproducts} for both FPMA and FPMB instruments.  \\

\section{Analysis \& Results}

\subsection{Timing analysis}

\begin{figure}
    \centering
    \includegraphics[scale=0.32,angle=0,trim={0cm 0.2cm 2cm 1cm},clip]{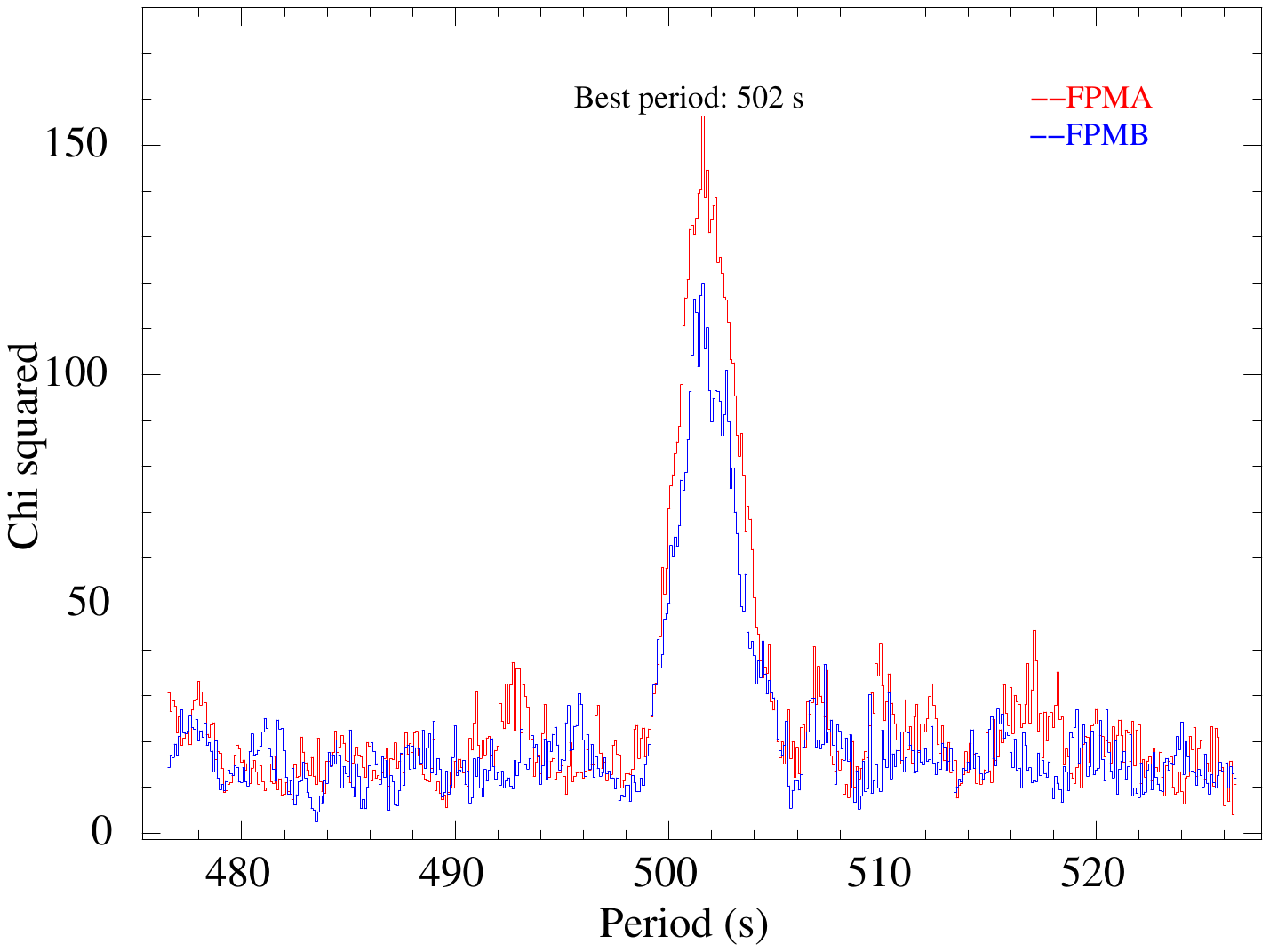}
    \caption{Period search results obtained using \code{efsearch} for \srcc ~shows a clear peak at 502~s in both detector modules. }. 
    \label{fig:efsearch}
\end{figure}

\begin{figure}
    \centering
    \includegraphics[scale=0.4,angle=0,trim={6cm 2.8cm 2cm 1cm},clip]{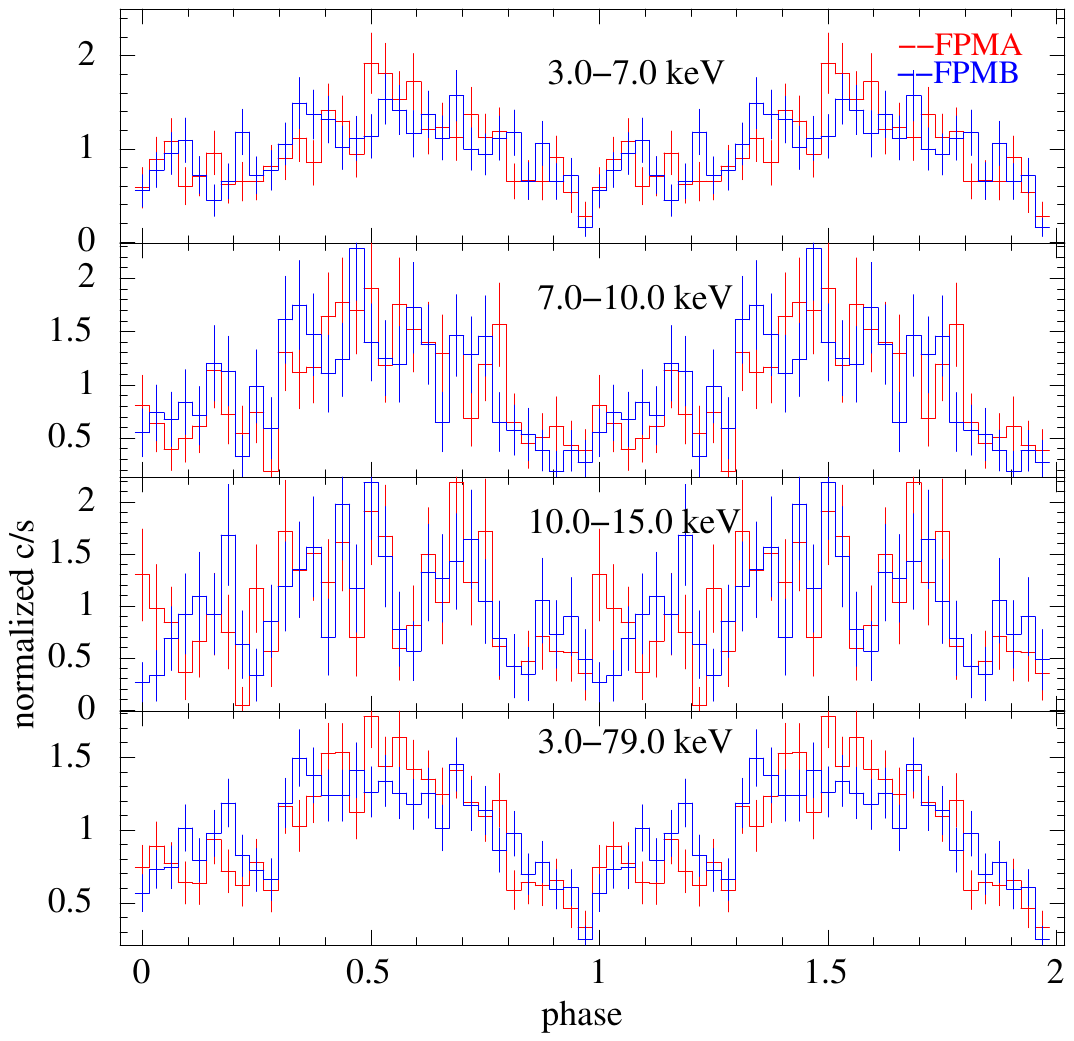}
    \caption{Energy resolved pulse profiles for \srcc ~light curves folded at the pulse period of 501.6 s. }. 
    \label{fig:efold}
\end{figure}

\begin{figure}
    \centering
    \includegraphics[scale=0.3,angle=0,trim={0cm 0.2cm 0cm 0cm},clip]{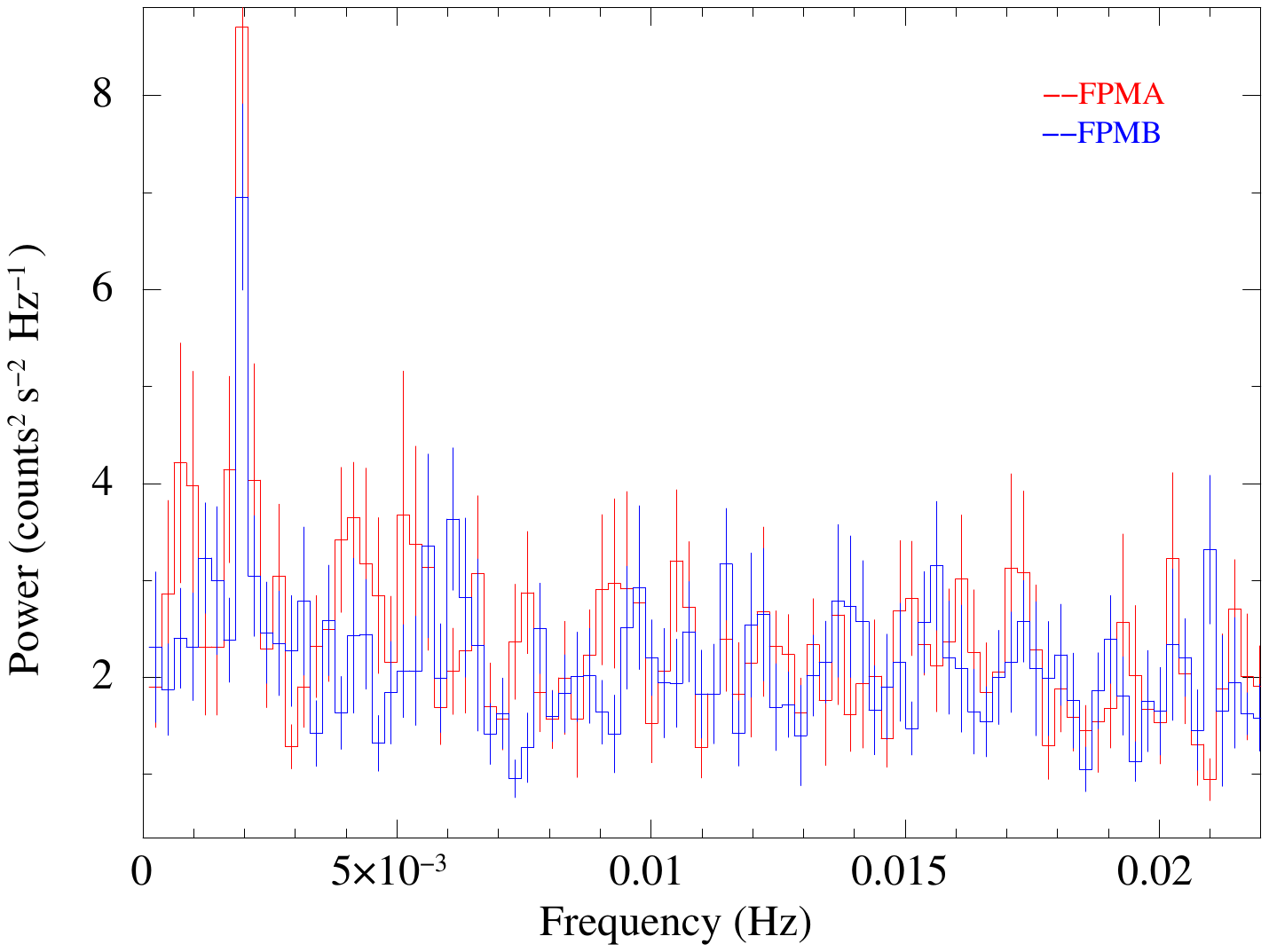}
    \caption{The power density spectrum for \srcc ~obtained using light curves from both detector modules showing the pulse peak at 1.99~mHz.}
    \label{fig:pds}
\end{figure}

In order to examine their low-level accretion properties, we carried out a pulsation search for the targets. The source and background light curves were generated from the FPMA and FPMB modules. The background-subtracted light curves have a mean count rate of $\sim$0.04 c~s$^{-1}$ and 0.06~c~s$^{-1}$, for \src ~and \srcc, respectively, in the 3--79~keV band. 
We further carried out a pulsation search using the \code{ftool} \code{efsearch} around the previously reported spin period values and also searched for any detectable QPO peaks in the power spectrum using the \code{FTOOLS} task - \code{powspec}. For \src, we were unable to detect any signatures of excess power at any frequency during this observation. Our results are consistent with previous null reports of pulsations during quiescence using \chandra~ \citep{Tsygankov2017quies}. 
In \srcc, we were able to identify a clear peak in \code{efsearch} at 502$\pm$3~s when the light curves from each of the detector modules were folded a resolution of 0.1 s (Figure \ref{fig:efsearch}). The measured period is consistent with previous measurements of 503$\pm$10~s during its 2010 outburst \citep{Kennea2010ATel2962a}. The uncertainty on the spin measurement is computed using a thousand realizations of Gaussian randomized light curves (a method similar to what is described in \citealt{Boldin2013} and \citealt{Varun2022}). Since \srcc ~was known to exhibit Quasi Periodic Oscillations (QPOs) during its low accretion state \citep{Kaur2010ATel3082,Donmez2020}, we carried out a search for QPOs in the power density spectrum between 0.0001 Hz to 1 Hz using a light curve binned at 0.5~s. The power density spectrum was generated by averaging out 16 individual segments with 8192 s duration each. In Figure \ref{fig:pds}, we show the PDS in the 0.0--4mHz frequency range, where we detect the pulse peak at 1.99~mHz with an SNR of $\sim$5$\sigma$ (Figure \ref{fig:pds}). However, we do not observe any features that resemble a QPO beyond 0.1~Hz (as observed in prior works such as \citealt{Donmez2020}) from both the detector modules in \srcc. The 3--79~keV light curve was folded at the observed pulse period of 502~s at an epoch of MJD 59828 and 32 folded phase bins. The pulse profile consists of a double-peaked shape with a pulse fraction\footnote{PF=(Imax - Imin)/(Imax + Imin), where Imax and Imin are the maximum and minimum normalized count rates associated with the pulse profile.} of 66\% in the entire \nustar ~band (Figure \ref{fig:efold} last panel). We further extracted light curves in different energy bands - 3--5~keV, 5--10~keV, and 10--15~keV. Due to high background beyond 15~keV, we were unable to detect pulsations or measure pulsed fractions. We folded the energy resolved light curves using the same fold parameters to obtain energy resolved pulse profiles as shown in Figure \ref{fig:efold}.  We obtain pulsed fractions of 74\%, 84\% and 88\% in the 3--5~keV, 5--10~keV, and 10--15~keV bands.

\subsection{Spectroscopy}
We further carried out a phase averaged spectral analysis for \src ~and \srcc ~in \code{XSPEC v.12.12.0}. For each individual pulsar, we carried out a joint fit using the FPMA and FPMB modules. The spectra were grouped using the tool \code{grppha} in such a way that they contain at least 20 counts per bin for \src ~and 30 counts per bin for \srcc, respectively. A constant normalization was included between the spectra from the two modules to account for any variations in the effective areas or potential uncertainties in their calibrations. The constant was kept fixed at 1.0 for FPMA and was allowed to vary for FPMB. We used the Tuebingen-Boulder ISM absorption model `\code{TBabs}' model to describe the line-of-sight neutral Hydrogen column density (N$_H$) assuming WILMS abundances \citep{Wilms2000} and VERN cross sections \citep{Verner1996}. 

For characterizing the continuum spectra of accretion powered X-ray pulsars during the quiescence, their spectra have been traditionally described using several combinations of one or more phenomenological models, which include the following models: 1) a simple power law (with or without the addition of thermal blackbody components), 2) a neutron star atmosphere (NSA) model, which assumes emission from a magnetized neutron star \citep{Pavlov1995} and 3) a model comprising of two Comptonization components \citep{Tsygankov2019}, among others. \chandra~observations of quiescent pulsars such as GRO J1750-27 and V0332+53, have been well described using the NSA model \citep{Escorial2019GRO,Elshamouty2016}. The double compton model assumes comptonization by hot and cold electrons and has been recently observed to fit spectra for at least three quiescent pulsars including A0535+262, GX 304-1 and X-Persei \citep{Tsygankov2019} (see \citealt{Sokolova2021} for a detailed theoretical interpretation of the compton double hump model for low accretion level emission).  We describe the fitting process adopted for each of the pulsars using some of the models below.

\subsubsection{\src}
\begin{figure*}
    \centering

    \includegraphics[scale=0.32,angle=0,trim={1.5 1.2cm 2cm 2.5cm},clip]{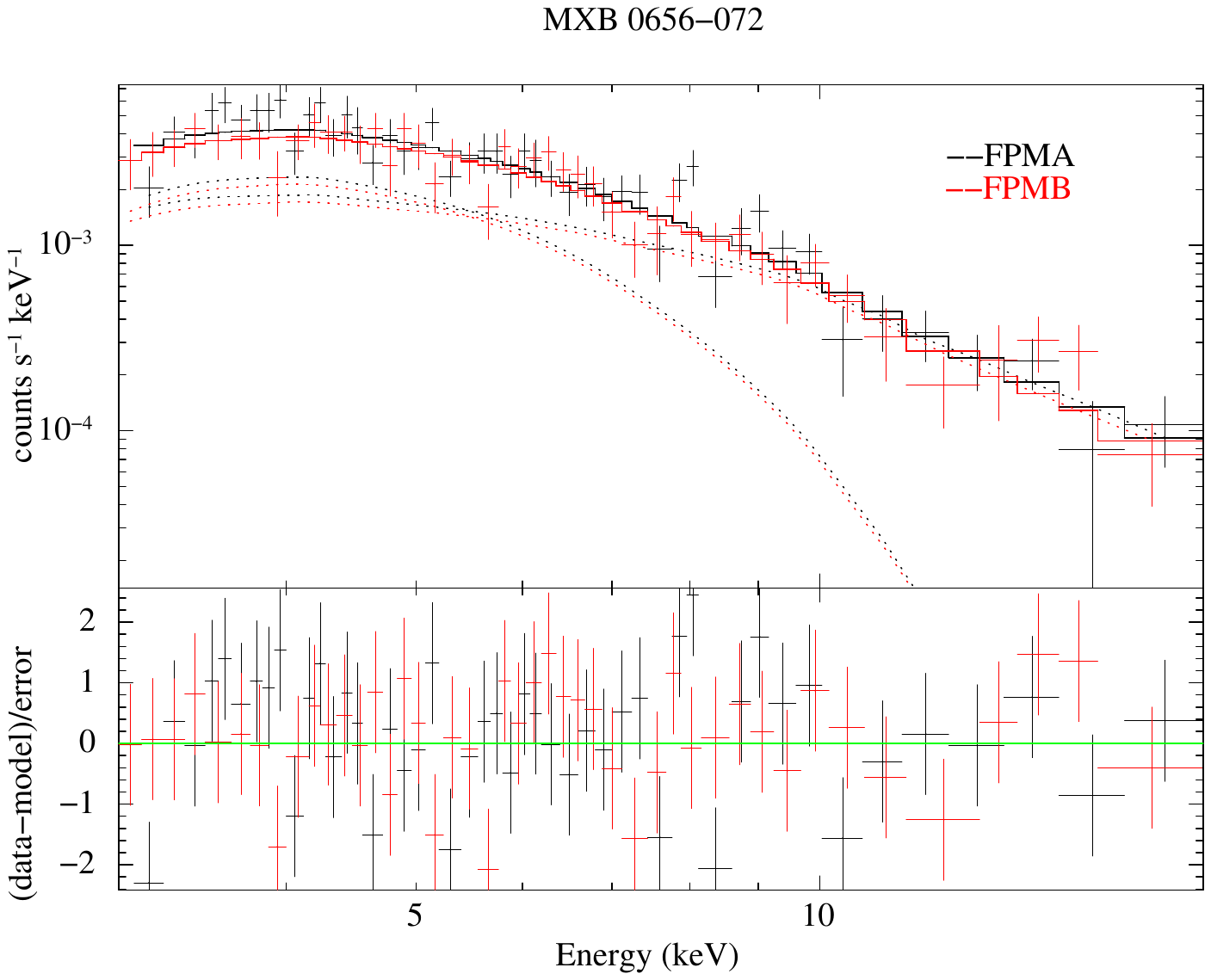}
    \includegraphics[scale=0.32,angle=0,trim={1 1.2cm 2.5cm 2.5cm},clip]{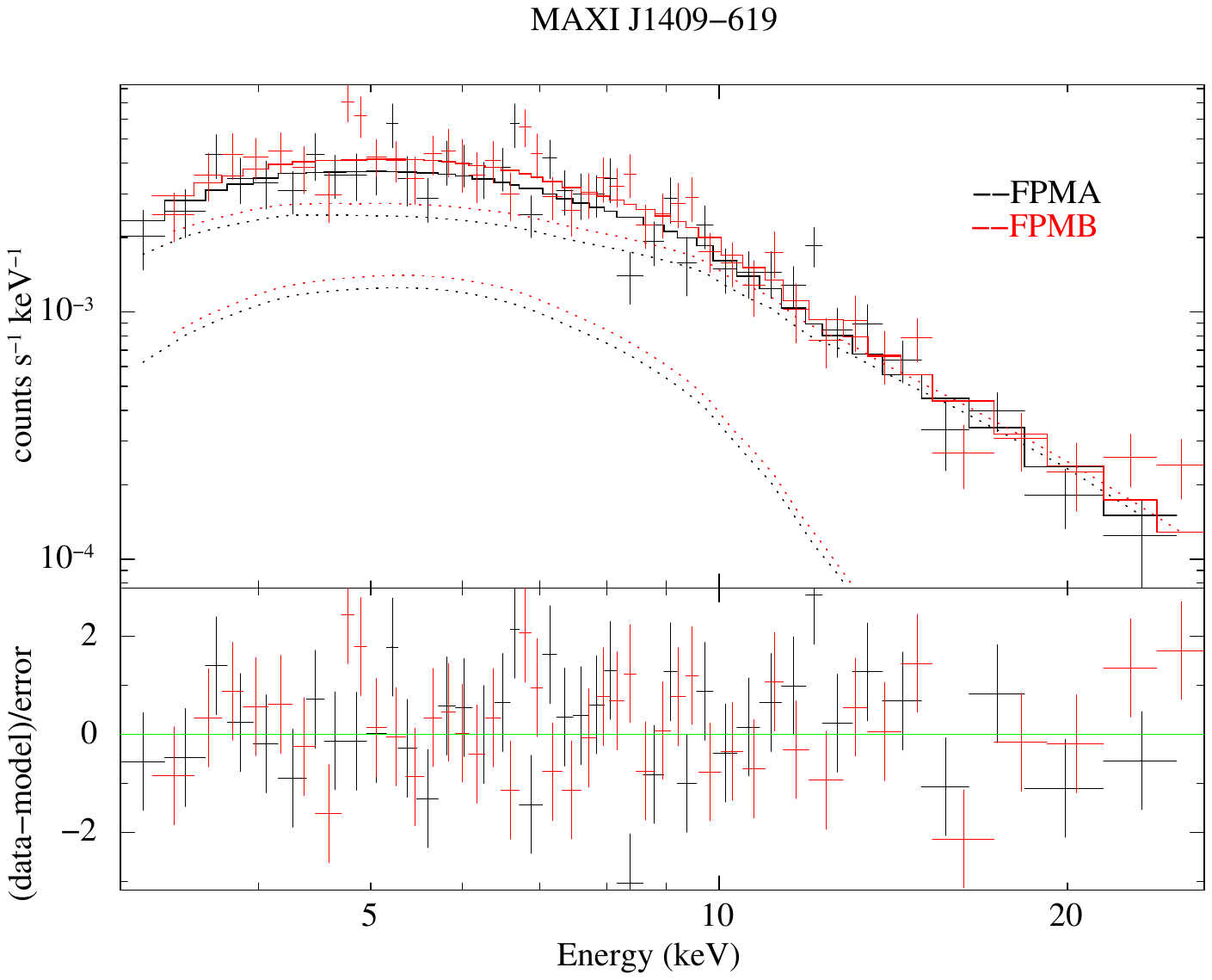}
    
    \caption{Best fit spectrum using the \code{PL+bbody} model for \src ~(left panel) and for \srcc ~(right panel).  The dotted lines indicate the additive \code{bbody} and \code{PL} components. For \src, the \code{PL} component dominates beyond $\sim$6~keV, while for \srcc ~the emission in the entire spectral band is dominated by the \code{PL} component. For both targets, the addition of the \code{bbody} model improved the fits.  } 
    \label{fig:specfig}
\end{figure*}

\src ~has been in a low luminosity state since its last outburst in 2008. Its quiescent state spectra, obtained using \chandra, four years after the outburst, had a thermal shape with a blackbody temperature around 1~keV and a luminosity of $\sim$4$\times$10$^{33}$~erg~s$^{-1}$ \citep{Tsygankov2017quies}. We use this as a starting point for the spectral analysis. Since the background counts start dominating above 20~keV, we ignore the energy range beyond 20~keV. The N$_\mathrm{H}$ was kept fixed at the galactic value\footnote{https://heasarc.gsfc.nasa.gov/cgi-bin/Tools/w3nh/w3nh.pl} of 0.6$\times$10$^{22}$~cm$^{-2}$ \citep{HI4PI2016}. We first fit the 3.0--20.0~keV spectrum using a simple absorbed power law model. This resulted in a reasonable fit with a $\chi^2$/dof of 93.2/93. We additionally also tried a combination of a blackbody and a powerlaw model that gave similar fit results. A single blackbody model was insufficient to fit the entire broadband spectrum, so we discard that model. We also alternatively attempted the fit using the partial covering absorber model (\code{pcfabs}) instead of the blackbody function. We found that the resultant blackbody fits were marginally better (reduced $\chi^2 \sim$0.91 for 93 dof) compared to the \code{pcfabs} (reduced $\chi^2$$\sim$0.96 for 107 dof). In addition to these models, we carried out a fit using a model comprising of two broad comptonized components as prescribed by \citet{Tsygankov2019}. We adopted the \code{compTT} model from \code{XSPEC} that describes the comptonization process of soft photons in a hot plasma \citep{Tit1994}. We tied the seed photon temperatures of both the components together and allowed the plasma temperature and plasma optical depth parameters to vary. The fit was unable to constrain the plasma temperature for both compton components. Alternatively, we tried modeling the possible two comptonized components using two power law functions. Although the fit did not seem to have any visible residuals, the resultant power law index values were not getting constrained. We therefore disregard the double Comptonization models for this paper. The fit parameters for all the models discussed above, except the double comptonization models (\code{comptt+comptt} and \code{po+po}), are shown in Table \ref{tab:specsrc1} and the spectral fit for the \code{PL+bbodyrad} model is shown in Figure \ref{fig:specfig} (left panel). The data statistics beyond 20~keV were poor to constrain the CRSF line parameters. 

In this current quiescent epoch, the source spectrum appears to be extremely soft. We used the \code{CFLUX} model to derive an average source flux of 1.5$\times$10$^{-12}$~ergs~cm$^{-2}$~s$^{-1}$ in the 3--79~keV band.  Assuming a source distance of 5.1~kpc \citep{Arnason2021}, we obtain a luminosity of 4.7$\times$10$^{33}$~erg~s$^{-1}$, which is lower by almost a factor of 2 from the previous \chandra~ measurement. 

\subsubsection{\srcc}

We have carried out the spectral fitting for \srcc ~following the spectral analysis from its previous low state observations using \sax \citep{Orlandini2012}. We fix the line of sight column density (N$_H$) to its galactic value of 2.0$\times$10$^{22}$~cm$^{-2}$. The grouped spectra were jointly fit for both the FPMA and the FPMB modules in \code{XSPEC} in the 3.0--30.0~keV band, since statistics were poor beyond that. We first fit an absorbed power law model which yielded a reasonable fit with a $\chi^2$/d.o.f of 107.4/86. Since the spectrum is background dominated above 30 keV, we were unable to identify signatures of the predicted CRSF line near 44~keV. To improve fit further and address the low energy residuals, we added a thermal blackbody component. This improved the fit ($\Delta \chi^2 \sim$ 9 for 86 dof). Alternatively, we also tried the fit using the partial covering absorber model (\code{pcfabs}). The resultant fits were marginally better with the blackbody model (reduced $\chi^2 \sim$1.10 for 86 dof) compared to using \code{pcfabs} (reduced $\chi^2$$\sim$1.26 for 90 dof). We again attempted to use a combination of two comptonization  models in the form of two \code{comptt} models and alternatively, two \code{PL} models. In both these cases, the fitting was unable to constrain the power law index or norm. Additionally, weird residuals at lower and higher energies showed up. We therefore ignore these double Comptonization models for further discussion. Details of the best fit parameters for the PL and PL+bbodyrad models are shown in Table \ref{tab:maxispec} and the spectral fit for the \code{PL+bbodyrad} model is shown in Figure \ref{fig:specfig} (right panel). \srcc ~had a source flux of 4.4$\times$10$^{-12}$~ergs~cm$^{-2}$~s$^{-1}$. Assuming a source distance of 14.5~kpc, we obtain a luminosity of 1$\times$10$^{35}$~erg~s$^{-1}$.

\begin{table}
    \centering
    
    \begin{tabular}{|c|c|c|}
         \hline
         Parameter/Model   &\code{PL} & \code{PL+BB}\\
         \hline
         \code{const} FPMA  &  1.0 (fixed) & 1.0 (fixed)\\
         \code{const} FPMB   & 0.97 & 0.97\\
         \code N$_{H}$ ($\times$10$^{22}$cm$^{-2}$)  &  0.6 (fixed) & 0.6 (fixed)\\
         $\Gamma$  & 2.69$\pm$0.11 & 2.1$^{+0.5}_{-0.7}$\\
         norm ($\times$10$^{-4}$) & & 5.6\\
         T$_\code{bbody}$ (\code{K})  & --  & 0.99$^{+0.16}_{-0.11}$\\
         R$_\code{bbody}$ (km)  & -- & 0.11$\pm$0.04  \\ 
         \hline
         Red. $\chi^2$(dof)  & 1.01(93) & 0.91(93)\\
         \hline
         Unabsorbed flux (3--79~keV)$^{1}$  & 1.5$\pm$0.15  & 1.82$\pm$0.2\\
         Unabsorbed flux (3--20~keV)$^{1}$  & 1.3$\pm$0.13  & 1.28$\pm$0.12\\
         
         
         \hline
    \end{tabular}
    \caption{Best fit spectral parameters for the joint FPMA+FPMB fits for \src ~assuming two model combinations as detailed in the text (see Section 3.2). Errors are indicated at 90\% confidence.\\  $^1$ In units of 10$^{-12}$~erg~cm$^{-2}$~s$^{-1}$} 
    \label{tab:specsrc1}
\end{table}


\begin{table}
    \centering
    
    \begin{tabular}{|c|c|c|}
         \hline
         Parameter/Model  & \code{PL} & \code{PL+BB}\\
         \hline
         \code{const} FPMA  &  1.0 (fixed) & 1.0 (fixed)\\
         \code{const} FPMB   & 1.1 & 1.1\\
         \code N$_{H}$ ($\times$10$^{22}$cm$^{-2}$)  & 2.0 (fixed) & 2.0 (fixed)\\
         $\Gamma$  & 1.73$\pm$0.07 & 1.54$\pm$0.25\\
         norm ($\times$10$^{-4}$) & 4.2 & 2.8\\
         T$_\code{bbody}$ (\code{K})  &   & 1.75$^{+0.8}_{-0.3}$\\
         R$_\code{bbody}$ (km) &  & 0.10$\pm$0.05$^{2}$  \\ 
         \hline
         Red. $\chi^2$(dof)   & 1.24(86) & 1.1 (86)\\
         \hline
         Unabsorbed flux (3--79~keV)$^{1}$   &  4.41$\pm$0.04  & 4.50$\pm$0.11\\
         Unabsorbed flux (3--30~keV)$^{1}$   &  2.70$\pm$0.02  & 2.63$\pm$0.03\\
         
         
         \hline
    \end{tabular}
    \caption{Best fit spectral parameters for the combined (FPMA+FPMB) spectra for \srcc ~assuming two model combinations as detailed in the text (see Section 3.2). Errors are indicated at 90\% confidence.\\ $^1$ In units of 10$^{-12}$~erg~cm$^{-2}$~s$^{-1}$.\\ $^2$ Assuming 10\% uncertainty in distance measurement for \srcc.} 
    \label{tab:maxispec}
\end{table}

\section{Discussion}

In this paper, we present the results from the analysis of \src ~and \srcc ~during one of their lowest-ever observed luminosity states using broadband \nustar ~observations conducted after a decade of quiescence. For the first time in these two systems, we have uncovered a dominant non-thermal contribution to the quiescent emission, whose origin we discuss below. Our results also reveal the presence of thermal hotspots that are indicative of ongoing accretion albeit at an extremely low level. In \srcc, we detect pulsed emission with a spin period of 502~s. 

\subsection{The propeller regime}
In the `propeller' regime \citep{IllSun1975}, a centrifugal barrier is generated by the rotating NS magnetosphere, due to which matter infall is halted. Further accretion will continue to remain inhibited if the velocity of the magnetic field lines is greater than the local keplerian velocity. Only when the magnetospheric radius shrinks below the corotation radius, which is easily realized during high accretion rates, can matter penetrate into the magnetosphere and can get accreted. Since the magnetospheric radius is a function of the mass accretion rate, one can determine the limiting luminosity for the onset of the propeller regime \citep{Campana2002,Tsygankov2016} by equating the magnetospheric radius and the corotating radius:

\begin{equation}
    L_{lim}(R) = \frac{GM{\dot{M}_{lim}}}{R} = 4\times10^{37}~k^{7/2}~B^{2}_{12}~P^{-7/3}~M^{-2/3}_{1.4}~R^{5}_{6} ~ \rm erg~s^{-1}
\end{equation}
Here, $M_{1.4}$ is the neutron star mass in units of $1.4~M_\odot$, $\dot{M}$ is the mass accretion rate, $R_6$ is the radius of the NS in units of 10$^{6}$~cm, $B_{12}$ is the magnetic field in units of 10$^{12}$~G and $P$ is the spin period of the pulsar in seconds. The factor $k$ is defined such that it relates the magnetospheric radius for disc accretion and the Alfv\`en radius for spherical accretion and is typically assumed to be 0.5 \citep{GhoshLamb1978,Tsygankov2016}. In several cases, accreting pulsars transitioning to the propeller regime have been associated with sudden luminosity drops (for example, 4U 0115+63 and V0332+53 \citealt{Campana2001,Tsygankov2016}). 

\src ~has been observed using \nustar at a luminosity of 4.7$\times$10$^{33}$~ergs~s$^{-1}$ which is lower from the previous quiescent state measurement using \chandra\ \citep{Tsygankov2017quies}. We assume canonical NS mass and radius values of 1.4~M$_{\odot}$ and 12~km \citep{Nattila2017}, respectively, and further input the measured spin period of 160.4~s \citep{McBride2006} and a magnetic field strength of 3.6~$\times$10$^{12}$~G \citep{Heindl2003}. We find the limiting luminosity for this source to be at 6.5$\times$10$^{32}$~erg~s$^{-1}$, which is down by a factor of 10 from the current source luminosity. For the case of \srcc, the observed luminosity during the recent \nustar ~observations is 1$\times$10$^{35}$~erg~s$^{-1}$. By using the spin period measured in this work (501.6~s) and the magnetic field estimate of 3.8$\times$10$^{12}$~G \citep{Orlandini2012}, we obtain a propeller limiting luminosity of 5.1$\times$10$^{31}$~erg~s$^{-1}$, which is well below the observed luminosity. Our findings are therefore in line with the expectations that both targets are in a low-accretion regime rather than a propeller regime.

\subsection{Origin of low-level emission}

Following the above discussion on the limiting luminosity for transition to the propeller regime, we can safely assume that during the \nustar ~observations for these two objects, accretion is not centrifugally inhibited in any way, which suggests that the observed X-ray emission is indicative of ongoing accretion. Here, we first discuss various quiescent emission mechanisms and compute their predicted luminosity estimates.  We will then discuss the applicability of these models for the two targets.

\subsubsection{The cold disk (CD) model}
Recent observations of slowly rotating pulsars accreting at low levels have demonstrated that not all pulsars undergo such transitions to the propeller regime \citep{Tsygankov2017cold}. It has been proposed that for slow pulsars, a slowly rotating magnetosphere can further push down the limiting luminosity, while maintaining temperatures $\sim$6500~K that would sustain recombination of hydrogen. This model is known as the `cold' accretion disk model \citep{Tsygankov2017cold,Tsygankov2017quies}. Below a critical mass  accretion rate ($\dot{M}_{cold}$), such recombinations are predicted to set in \citep{Lasota1997,Tsygankov2017cold}: 
\begin{equation}
  \dot{M}_{cold}=3.5\times10^{15}~r_{10}^{2.65}~M_{1.4}^{-0.88} \rm g~s^{-1}  
\end{equation}
where $r_{10}=r/10^{10}$~cm is the inner disc radius. The luminosity at which the transition to a cold disc occurs will correspond to the following:
\begin{equation}
     L_{cold}=9\times10^{33}~k^{1.5}~M_{1.4}^{0.28}~R_{6}^{1.57}~B_{12}^{0.86}~\rm erg~s^{-1}
\end{equation}

This is $\sim$2$\times$10$^{34}$~erg~s$^{-1}$ for typical pulsars with magnetic field strengths of $\sim$10$^{12}$~G. Here the value of $k$ is taken as 0.5 as commonly considered for disc accretion \citep{GhoshLamb1978}. If $L_{\rm cold}>L_{\rm prop}$, then lower mass accretion rates can drive such systems to stably accrete from a cold disc. In the standard accretion scenario, the magnetospheric radius increases as the source gets dimmer. As opposed to that, for pulsars that transition to the `cold' phase, the inner disk radius decreases at lower luminosity states. This ensures that these sources continue to stably accrete from a cold disc before switching to the propeller regime at even lower luminosities \citep{Tsygankov2017quies}. Furthermore, according to this model, the luminosity is expected to fade as $L\propto t^{-0.7}$ and observations of sustained pulsed emission can be expected \citep{Tsygankov2017cold}. The transition luminosity to the cold disc regime and the corresponding predicted luminosities (expected to be observed at any given $t$) are shown in Table \ref{tab:lum}. 

\subsubsection{The deep crustal heating (DCH) model}
An alternative explanation for the origin of quiescent X-ray emission in pulsars is the deep crustal heating model. According to this model, the soft thermal component is assumed to arise from a cooling neutron star which is powered by nuclear reactions that occur during each outburst \citep{Escorial2020,Brown1998}. Matter that is deposited on the NS surface during every outburst compresses the deeper crustal layers, which then triggers nuclear reactions and releases heat. This will drive the heated crust out of equilibrium with the NS core. After the outburst is complete, and quiescence sets in, thermal radiation is emitted in the form of cooling radiation from the NS surface and eventually brings the crust and core back into equilibrium (see \citealt{Escorial2019GRO}, \citealt{WijDegenaar2017} and section 4.3 in \citealt{Tsygankov2017quies} for a detailed review). 


For a population of non-pulsating LMXBs and accreting ms pulsars \citep{WijDegenaar2017}, it appears that enhanced cooling processes such as the Direct URCA process (as against standard neutrino cooling via the modified URCA or bremsstrahlung processes) may need to be invoked in order to explain the observed crustal cooling curves. \citet{WijDegenaar2017} elaborate on the various cooling mechanisms applicable for different particle compositions of the NS core. In addition to the DCH model, a `shallow heating' mechanism has also been invoked in the past in order to explain crust cooling curves seen in LMXBs \citep{Escorial2019GRO, Deibel2015}. Be X-ray pulsars being relatively young systems, may not possess a crust that is completely made of accreted material. Instead, they can be expected to have as `hybrid' crusts \citep{WijDegenaar2017}. It is unclear whether such hybrid crusts can support all kinds of accretion induced nuclear reactions \citep{WijDegenaar2017} or whether some of the deep crustal reactions that generate the necessary heating get inhibited \citep{Escorial2019GRO}. Moreover, the presence of `shallow heating' effects remains largely explored for Be pulsars (see \citealt{Escorial2019GRO} for the singular case of GRO J1750-27), since the effect of strong magnetic fields on the heating and cooling of the NS crusts is barely understood. For the Be systems that \textit{have} shown signatures of cooling of an accretion heated NS, the emission has been found to arise from a small portion confined to the NS hotspots \citep{Campana2002,Elshamouty2016,WijDegenaar2017}. 4U 0115+63 and V0332+53 are the only two Be pulsars, for which signatures of crustal heating and cooling have been identified \citep{WijDeg2016,Escorial2017}. Additional searches for signatures of crust cooling were carried out in other Be pulsars such as GRO 10750-27 \citep{Escorial2019GRO}, with no success. 

\begin{figure}
    \centering
    \includegraphics[scale=0.26]{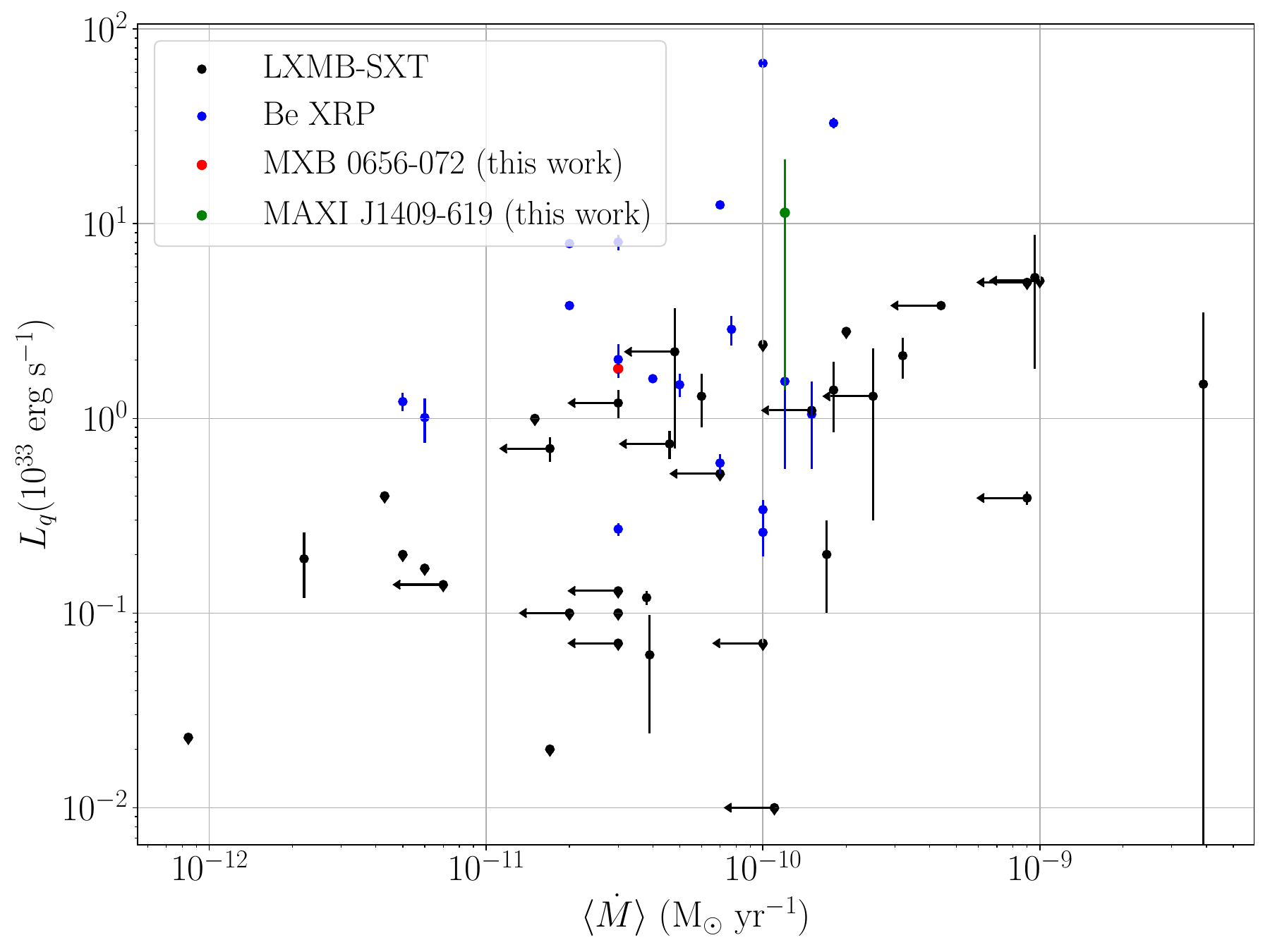} 
    \caption{Plot showing the measured quiescent thermal luminosity as a function of the long term averaged mass accretion rate $\langle\dot{M}\rangle$. The quiescent luminosity data for LMXBs have been plotted from \citep{Potekhin2019} and those for Be X-ray pulsars are adopted from \citep{Tsygankov2017quies}. Results for \src ~and \srcc ~are marked in red and green, respectively. We assume a 10\% uncertainty on the 14.5~kpc distance measurement for \srcc ~since no previously reported uncertainty exists in the literature.}
    \label{fig:qlummacc}
\end{figure}

In order to understand the thermal heating and crust cooling effects in all categories of accreting NS systems, we construct a comprehensive plot of the observed quiescent luminosities as a function of their long term mass accretion rate (Figure \ref{fig:qlummacc}). This plot includes compilations of measured quiescent thermal luminosities of 1) LXMBs, accreting ms-pulsars and Soft X-ray Transients (SXT) from \citet{Potekhin2019}, and 2) of Be-X-ray pulsars from \citet{Tsygankov2017quies}. Several key factors play into the estimation of the quiescent thermal luminosity for accreting neutron stars. These include the long term averaged mass aceretion rate, the composition of the crust, and the most relevant cooling process. We note that several uncertainties exist within these measurements. These include variations in the long term average accretion rate (during different quiescent epochs), measurement methods of the long term accretion rate, and distance estimates, to name a few. We therefore treat these estimates as crude. This plot can serve as a reference to test various crust cooling scenarios (as shown in \citealt{WijDegenaar2017}).

In order to derive the expected quiescent luminosities from deep crustal heating, we adopt a procedure similar to what has been indicated in \citet{Tsygankov2017quies} and \citet{Elshamouty2016}. We use the long term \swift-BAT light curve count rates and convert it to flux using WebPIMMS\footnote{https://heasarc.gsfc.nasa.gov/cgi-bin/Tools/w3pimms/w3pimms.pl} and WebSPEC\footnote{https://heasarc.gsfc.nasa.gov/webspec/webspec.html} by assuming a simple power law model with an index 1.0 and a high energy cutoff at 15~keV, which is a representative spectrum for typical X-ray pulsars in this energy band. We use this flux to then estimate the average long term luminosities for both targets. Assuming perfect accretion efficiency, we then compute the mass accretion rate as follows:
\begin{equation}
 \langle\dot{M}\rangle=\frac{L_{avg}~R}{GM} 
\end{equation}
where $G$ is the gravitational constant, $M$ is the mass and $R$ is the radius of the neutron star. We assume canonical neutron star mass and radius values of 1.4~M$\odot$ and 12~km, respectively.  This gives us an accretion rate of 0.3$\times$10$^{-10}$~M$\odot$~yr$^{-1}$ and 1.1$\times$10$^{-10}$~M$\odot$~yr$^{-1}$, for \src ~and \srcc, respectively. We then  estimate the expected quiescent luminosity as described in \citet{Tsygankov2017quies} and \citet{Brown1998} as follows:
\begin{equation}
     L_{q,predicted, DCH}=\frac{ \langle  \dot{M} \rangle}{10^{-11}~M_{\odot}~\rm yr^{-1}}\times6\times10^{32} ~\rm erg~s^{-1}
\end{equation}
This gives us predicted quiescent state thermal luminosities of 2.0$\times$10$^{33}$~erg~s$^{-1}$ and 1$\times$10$^{34}$~erg~s$^{-1}$, for \src ~and \srcc, respectively. The predicted luminosities are shown in Table \ref{tab:lum}.

\begin{table*}
    \centering
    \begin{tabular}{|c|c|c|c|c|c|c|c|c|}
         \hline
        Source & Period$^{1}$ & L$_{\rm prop}$ & $\langle\dot{ 
        \rm M}\rangle$ & L$_{\rm q, predicted, DCH}$  & L$_{\rm q, transition~ to~CD}$ & L$_{\rm q, predicted, CD}$ & L$_{\rm bb}$ & L$_{\rm PL}$  \\
        & (s) & (10$^{32}$ erg~s$^{-1}$) & (10$^{-10}$~M$\odot$~yr$^{-1}$) & (10$^{33}$ erg~s$^{-1}$) & (10$^{34}$ erg~s$^{-1}$)  & (10$^{31}$ erg~s$^{-1}$)  &  (erg~s$^{-1}$) & (erg~s$^{-1}$)  \\
        \hline
         \src & - & 6.5 & 0.3 & 2.0  & 1.2 & 0.5 & 1.6$\times$10$^{33}$ & 4.1$\times$10$^{33}$ \\
         \srcc & 502 & 0.5 & 1.1 & 10.0 &  1.3 & 1.9 & 1.1$\times$10$^{34}$ & 10.4$\times$10$^{34}$ \\
         \hline
    \end{tabular}
    \caption{Parameters determined in this study for \src ~and \srcc. \\ 
    $^1$ Period measured during quiescence.}
    \label{tab:lum}
\end{table*}

\subsubsection{Contribution from the companion star}
Some fraction of the observed low-level X-ray emission can potentially be expected to come from the companion star. The companion for \src ~is a Be star and that for \srcc ~has not been confirmed as yet. However, detailed estimates carried out by \citet{Tsygankov2017quies} indicate that the X-ray luminosities of typical O or B type stars are of the order of $\sim$10$^{31}$~ergs~s$^{-1}$, which is several orders of magnitude lower than our measured luminosities for both \src ~and \srcc ~as part of this work. We therefore assume that the companion does not, in any significant way, contribute to the observed X-ray emission for both sources. 

\subsubsection{Likely emission mechanism in \src ~and \srcc}

The Be X-ray source \src ~continues to exhibit thermal emission consistent with the predicted emission expected from the cooling of accretion heated neutron star during the current \nustar observation (this work and \chandra\ observations analyzed in \citealt{Tsygankov2017quies}). The radius of the blackbody emitting region ($\sim$0.06~km) is consistent with that of a NS hotspot near the poles. In addition, we have detected a strong power law component, for the first time, dominating over the thermal flux (see table \ref{tab:lum}), as against a purely thermal spectrum observed using \chandra\ \citep{Tsygankov2017quies} in 2012. The origin of this power law component provides additional incentive to better understand the processes occurring at low luminosities (see Section \ref{sec:originpl} below for detailed discussion). The cold accretion disk model (which supports continued accretion and therefore, pulsations) predicts a quiescent luminosity of 1.3$\times$10$^{31}$~erg~s$^{-1}$, which is two orders of magnitudes lower than what we observe using \nustar. This makes it an unlikely scenario for the current observations. Being a slowly rotating pulsar, it is unlikely that it will ever descend into a propeller regime where accretion is centrifugally inhibited. Our inability to detect pulsations for this pulsar using \nustar ~possibly hints at very low pulse fraction during this low state observation. On the other hand, we detect the presence of a thermal component (which is preferred over a partial covering absorber) from our spectral fits, which we can attribute to that arising from deep crustal heating. Our measured quiescent luminosity obtained from the thermal component for \src ~corresponds to 1.6$\times$10$^{33}$~erg~s$^{-1}$. The predicted quiescent state luminosities due to the deep crustal heating model is $\sim$2.0$\times$10$^{33}$~erg~s$^{-1}$, which is of the order of the observed thermal luminosities and is therefore likely to be the dominant source of quiescent emission in \src.

\begin{figure}
    \centering
    \includegraphics[scale=0.28,trim={0cm 0cm 0.2cm 0.3cm},clip]{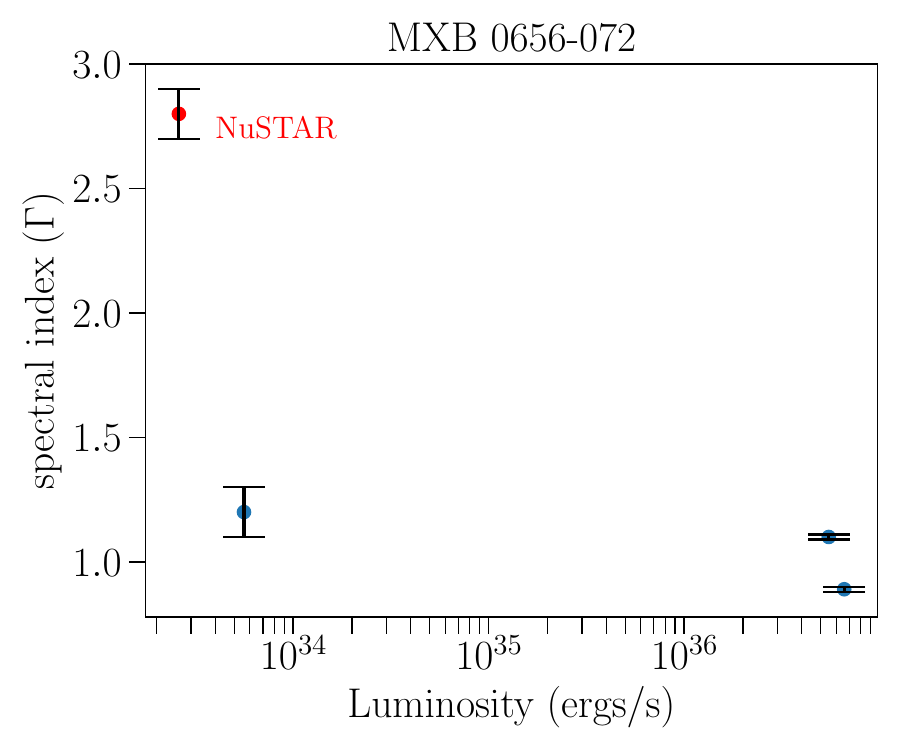}
    \includegraphics[scale=0.28,trim={0cm 0cm 0.2cm 0.3cm},clip]{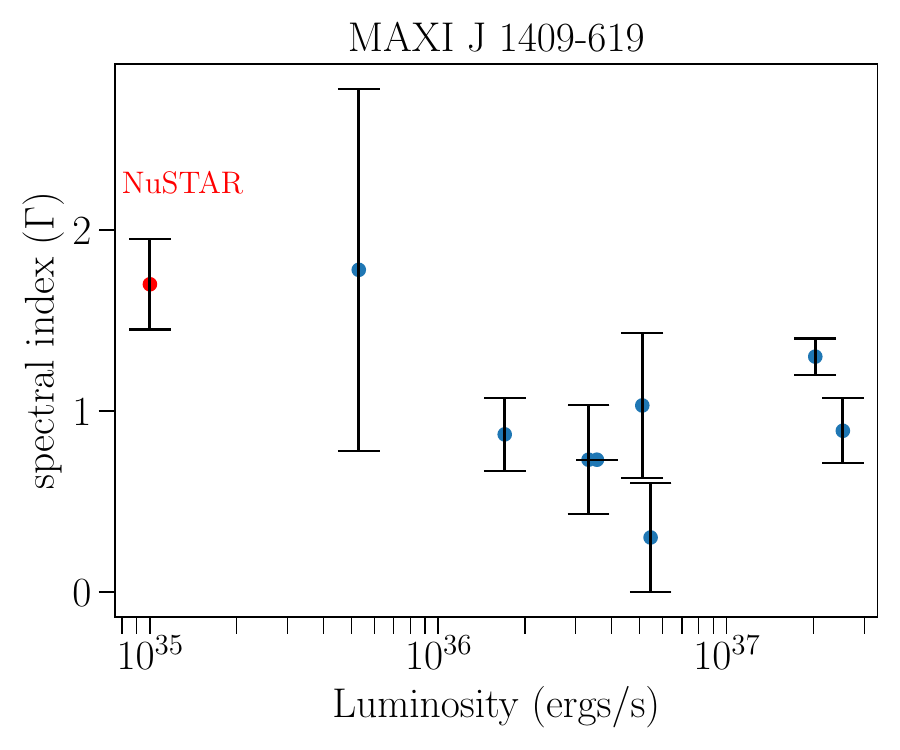}
    \caption{Variation of the spectral index as a function of luminosity. The red data point indicates the \nustar ~measurement obtained from this work.}
    \label{fig:specindexlum}
\end{figure}

In \srcc, the quiescent emission is observed to be extremely soft ($\Gamma \sim$1.6). Interestingly this source is among a few others that exhibit positive flux-spectral index correlations beyond 10$^{-9}$~erg~cm$^{-2}$~s$^{-1}$ (for example EXO 2030+275, \citealt{Epili2017}) and anti-correlation below these fluxes \citep{Donmez2020}. Such correlations and anti-correlations are also observed in sources that exhibit cyclotron features \citep{Malacaria2015,Donmez2020}. Our current low luminosity \nustar ~observations align with this anti-correlation trend (Figure \ref{fig:specindexlum}). We also detect strong pulsations at 502~s which indicates ongoing low-level accretion. The decade-long quiescence should have ensured that the NS crust and core are at thermal equilibrium. The predicted luminosity for the transition to a cold disk is for \srcc ~is estimated as $\sim$10$^{34}$~erg~s$^{-1}$. This would mean that during our current observations, \srcc ~could not have transitioned to this state. Moreover, the predicted luminosity as would be observable during these observations in September 2022 is of the order of 1.9$\times$10$^{31}$~erg~s$^{-1}$ (assuming an L $\propto$ t$^{-0.7}$ behavior, \citealt{Tsygankov2017cold}). This is way lower than the observed luminosity using \nustar. We disregard the cold accretion disk model for \srcc ~as well. 
The observed residual thermal emission likely arises from the NS hotspots which were heated during outbursts. This is further supported by the size of the emission region, which is 0.1~km (far less than the NS radius). Thermal emission from accretion-heated hot spots has similarly been observed for other Be pulsars such as 4U 0115+63 \citep{Escorial2017} and V0332+53 \citep{Elshamouty2016}. Keeping in mind the uncertainties in the distance estimates for \srcc, the measured thermal luminosity from this work seems to prefer the DCH model over the CD model.

\subsection{Detection of pulsation in \srcc ~at low accretion states}

At luminosities $\sim$10$^{36}$~erg~s$^{-1}$, \citet{Orlandini2012} reported detection of CRSFs in \srcc ~(the fundamental and its two harmonics) in the \sax spectrum. However, no pulsations were reported from those observations in 2000. Our results are indicative that pulsed emission can be observed at low accretion levels as well. Indeed pulsations have been detected in several low luminosity Be pulsars - for example, SAX J2103.5+4545 \citep{Reig2014}, 4U 078-25, GRO J1008-57 \citep{Tsygankov2017cold}, 2S 1417-624 \citep{Tsygankov2017quies}, etc. In some highly magnetized systems, only the hard X-ray component was found to be pulsed (see Cep X-4, \citealt{McBride2007} and RX J0812.4–3114 \citealt{Zhao2019}). \srcc ~was observed at a luminosity of $\sim$10$^{35}$~erg~s$^{-1}$ during the current \nustar ~observations, which happens to be one of its lowest to date, though not low enough to drive the system to a propeller regime. A strong pulsation has been measured at 502~s. Table \ref{tab:period} indicates all previously measured spin values using \swift \citep{Kennea2010ATelb}, \rxte \citep{Donmez2020} and the \fermi-GBM \citep{Malacaria2020} for \srcc. 

\begin{table}
    \centering
    \begin{tabular}{|c|c|c|}
    \hline
        Mission & MJD & Pulse period (Hz)  \\
        \hline
        Swift XRT$^1$ & 55530 & 0.001988$\pm$3.9524e-5 \\
       Fermi-GBM$^2$  & 55531.96 & 0.00197045$\pm$9.14e-8  \\
       Fermi-GBM$^2$  & 55535.97 & 0.00197707$\pm$7.61e-8 \\
       Fermi-GBM$^2$  & 55540.01 & 0.00198282$\pm$8.99e-8 \\
       RXTE-PCA$^3$ & 55540.0150 & 0.00198287$\pm$0.0004\\
       Fermi-GBM$^2$  & 55544.02 & 0.00198792$\pm$1.40e-7 \\
       Fermi-GBM$^2$  & 55548.09 & 0.001992423$\pm$1.34e-7\\
       
        NuSTAR$^4$ & 59828.95 & 0.00199203$\pm$1.19e-5 \\
        \hline
    \end{tabular}
    \caption{Table of measured pulse period for \srcc ~since the time of its discovery in 2010.\\
    $^1$\citet{Kennea2010ATelb} \\
   $^2$ \citet{Malacaria2020}\\
   $^3$ \citet{Donmez2020}\\
     $^4$ This work. }
    \label{tab:period}
\end{table}

Several X-ray pulsars have exhibited a large pulsed fraction during quiescence. For example, in RX J0812.4–3114, the hard component was observed to be pulsed with a pulsed fraction of $\sim$88\% \citep{Zhao2019}. The majority of pulsars studied by \citet{Tsygankov2017quies}, which had detectable pulsed emission (refer to Table 3 in \citealt{Tsygankov2017quies}), displayed significant pulsed fractions within the 50\% to 70\% range. High pulsed fractions have been found to be common for X-ray pulsars at low accretion states \citep{LutTsy2009} (similar to this work for \srcc) as well as for pulsars in complete quiescence \citep{Negueruela2000, Reig2014}. Variations in the behavior of pulsed fractions as a function of energy as well as luminosity states have been characterized for several X-ray pulsars in terms of a simple accretion model that is a function of the emission region and accretion geometry \citep{LutTsy2009}.  

It is interesting to note here that we do not yet know the nature of the companion star in \srcc. The source underwent an outburst in 2010 and has been in quiescence since. Based on what appears to be the erratic outburst of the source, it can be compared to a class of X-ray transients called the Supergiant Fast X-ray Transients (SFXTs), like IGR J17544-2619, rather than persistent supergiant X-ray binaries like Vela X-1. SFXTs exhibit similar properties as classical systems, including supergiant companions and orbital period distribution. However, SFXTs display greater dynamic variability than classical systems, characterized by sporadic short X-ray outbursts and faint flares with fast rise times (tens of minutes) and typical durations of a few hours. On average, SFXTs have X-ray luminosities 2-3 orders of magnitude lower than classical systems with similar orbital periods, outside of these outburst events \citep{walter2015}. There have been debates in the literature about the nature of compact objects in SFXTs. Should \srcc ~be classified as an SFXT, the presence of a pulsating NS would be in favor of a NS compact object scenario. In addition, the presence of cyclotron line at 44\,keV could further indicate that magnetic field strengths of SFXTs can be $\sim$10$^{12}$~G, contradicting some leading theories about them being accreting magnetars \citep{bozzo2008}. 


\subsection{Non-thermal emission at low luminosities} \label{sec:originpl} 
In literature, there are at least three Be XRPs (A0535+262, GX 304-1 and X Persei) that have displayed an unusual spectrum at low luminosities - the presence of non-thermal components like the two comptonization components \citep{Tsygankov2019}. The detection of a CRSF during quiescence (for example, A0535+262, \citealt{Tsygankov2017quies}), also indicates that the broadband spectrum of Be XRPs in quiescence can be modeled with non-thermal components since CRSFs are generated through non-thermal origin (inverse Compton scattering of electrons off photons). Motivated by these findings, we modeled the broadband spectra of \src ~and \srcc ~using two non-thermal components (two comptonization and two powerlaw). Although we were not able to constrain both the non-thermal components - given the faintness of both sources compared against other bright Be XRPs like A0535+262 in quiescence - we could constrain at least one non-thermal component.  The detection of this non-thermal component is crucial as it indicates ongoing accretion even in quiescence. The power law component has been generally understood to arise due to low-level accretion in LMXBs \citep{WijnandsDeg2015lmxb} and even in some X-ray pulsars such as RX J0812.4-3114 \citep{Zhao2019}. In pulsars such as GS 0834-430,  Swift J1626.6-5156, Cep X-4, etc., it was proposed that the non-thermal emission arose as a result of low-level accretion from a cold disc \citep{Tsygankov2017quies}. For the case of \srcc, the cold disc model is not favored. Moreover, an accretion column is not expected to form at such low luminosity states. Additionally, the lack of knowledge about the companion star can indicate alternative sources of origin for the non-thermal emission. Possible avenues include the accretion disk (wind-fed systems could still form disks, \citealt{Karino2019}) or the companion star's circumstellar disk.

It is worthwhile to note that the presence of cyclotron lines at 33~keV and 44~keV for \src ~and \srcc, respectively, outside quiescence, also indicate the non-thermal origin of X-rays in these systems out of quiescence. Although we were limited by statistics to constrain these features in the quiescent spectrum, these features are tell-tale signatures of the non-thermal origin of X-rays in these systems. 

The study of such low luminosity HMXBs in quiescence can be greatly boosted with instruments having a larger effective area in hard X-rays. One such proposed probe class mission is the High-Energy X-ray Probe (HEX-P; \citealt{Madsen2019HEX}). HEX-P provides focused hard X-rays up to $\sim$ 150\,keV using two high-energy telescopes (HETs) and soft X-ray coverage with a low-energy telescope (LET) with a much larger area than \nustar. HEX-P's unique capabilities would enable the study of accretion onto neutron stars across a wide range of energies, including in low-luminosity regimes like the ones studied in this paper.

\section{Conclusions}
In this work, we study the quiescent state properties of \src ~and \srcc ~using sensitive \nustar ~observations. We detect a strong pulsation at 502~s in \srcc ~with a 66\% pulsed fraction in the folded pulse profile. Our measured spin period is consistent with previous measurements. The observed pulse profile is typical of X-ray pulsars accreting at low levels. The pulsed fraction is seen to have an increasing trend as a function of energy, which is also consistent with what is observed in other X-ray pulsars. We do not detect QPOs in the power density spectra for \srcc . 

The fluxes measured in this work indicate that neither of these two sources, \src ~and \srcc, is in the propeller regime despite observations being carried out luminosities of  2.6$\times$10$^{33}$~erg~s$^{-1}$ and 1$\times$10$^{35}$~erg~s$^{-1}$, respectively. We show that the broadband spectral fits for both pulsars are best described using a combination of thermal and non-thermal emission components. The non-thermal component, which is very soft, is likely to emerge as a result of low-level accretion. From the measured radius of the blackbody emitting region, we infer that the thermal emission likely arises from the accretion heated hot spots on the surface of the NS. The detection of pulsed emission along with non-thermal emission in \srcc ~is indicative of ongoing accretion during quiescence. Future sensitive broadband observations of X-ray pulsars during quiescence will be useful to further understand low luminosity accretion processes.

\section*{Acknowledgements}
This research has made use of data and software provided by the High Energy Astrophysics Science Archive Research Center (HEASARC), which is a service of the Astrophysics Science Division at NASA/GSFC and the High Energy Astrophysics Division of the Smithsonian Astrophysical Observatory. The analysis work has made use of the NuSTAR Data Analysis Software (NuSTARDAS) jointly developed by the ASI Space Science Data Center (SSDC, Italy) and the California Institute of Technology (Caltech, USA). G.R is supported by NASA under award number 80NSSC22K1814. G.R is grateful to Deepto Chakraborty and Ron Remillard for useful discussions. 


\section*{Data Availability}

The observational data utilized in this work is publicly available through the High Energy Astrophysics Science Archive Research Center (HEASARC). Any additional information will be shared on reasonable request to the corresponding author.




\bibliographystyle{mnras}
\bibliography{example} 








\bsp	
\label{lastpage}
\end{document}